\documentclass[aps,onecolumn,groupedaddress,nofootinbib,showpacs]{revtex4}
\usepackage{graphicx}
\usepackage{listings}
\usepackage{amsmath}
\usepackage{bm}

\newcommand{\bfrN}{{\bf r}^N}
\newcommand{\dbfrN}{{\rm d}{\bf r}^N\ }

\newcommand{\bfq}{{\bf q}}
\newcommand{\bfqT}{{\bf q}^T}
\newcommand{\bfqs}{{\bf q}^*}
\newcommand{\bfqsT}{({\bf q}^*)^T}
\newcommand{\dbfq}{{\rm d}{\bf q}}
\newcommand{\bfPsi}{\boldsymbol{\Psi}}
\newcommand{\Dpsi}{\Delta\psi}
\newcommand{\bpsi}{\bar{\psi}}

\newcommand{\matA}{\mathbf{A}}
\newcommand{\matAinv}{\mathbf{A}^{-1}}
\newcommand{\matS}{\mathbf{S}}
\newcommand{\bfBz}{{\bf B}(\bfrN)}
\newcommand{\bfD}{{\bf D}}
\newcommand{\bfDT}{{\bf D}^T}
\newcommand{\bfE}{{\bf E}}
\newcommand{\bfET}{{\bf E}^T}
\newcommand{\bfG}{{\bf G}}

\newcommand{\mathH}{\mathcal{H}}
\newcommand{\mathK}{\mathcal{K}}
\newcommand{\mathU}{\mathcal{U}}
\newcommand{\mathUz}{\mathcal{U}_0}

\newcommand{\mathZ}{\mathcal{Z}}
\newcommand{\mathP}{\mathcal{P}}
\newcommand{\mathp}{p}
\newcommand{\mathO}{\mathcal{O}}

\newcommand{\avgpsi}[1]{\left< #1 \right>_{\bfrN, \bfq}}
\newcommand{\avgBO}[1]{\left< #1 \right>^\mathrm{BO}_{\bfrN}}

\begin{document}

\title{Charge fluctuations from molecular simulations in the constant-potential ensemble} 

\author{Laura Scalfi,\textit{$^{a}$} 
David Limmer,\textit{$^{b,c,d,e}$} 
Alessandro Coretti,\textit{$^{f,g}$}
Sara Bonella,\textit{$^{g}$}
Paul A. Madden,\textit{$^{h}$}
Mathieu Salanne,\textit{$^{a,i}$}
and Benjamin Rotenberg$^{\ast}$\textit{$^{a,i}$}
}
\affiliation{
\textit{$^{a}$}~Sorbonne Universit\'{e}, CNRS, Physicochimie des \'Electrolytes et Nanosyst\`emes Interfaciaux, F-75005 Paris, France.\\
 \textit{$^{b}$}~Department of Chemistry, University of California, Berkeley, CA, USA.\\
 \textit{$^{c}$}~Kavli Energy NanoScience Institute, Berkeley, CA, USA.\\
 \textit{$^{d}$}~Materials Science Division, Lawrence Berkeley National Lab., Berkeley, CA, USA.\\
 \textit{$^{e}$}~Chemical Science Division, Lawrence Berkeley National Lab., Berkeley, CA, USA.\\
 \textit{$^{f}$}~Department of Mathematical Sciences, Politecnico di Torino, I-10129 Torino, Italy.\\
 \textit{$^{g}$}~Centre Europ\'een de Calcul Atomique et Mol\'eculaire (CECAM), Ecole Polytechnique F\'ed\'erale de Lausanne, 1015 Lausanne, Switzerland.\\
 \textit{$^{h}$}~Department of Materials, University of Oxford, Oxford, UK.\\
 \textit{$^{i}$}~R\'eseau sur le Stockage Electrochimique de l'Energie (RS2E), FR CNRS 3459, France.}

\begin{abstract} 
We revisit the statistical mechanics of charge fluctuations in capacitors. In constant-potential classical molecular simulations, the atomic charge of electrode atoms are treated as additional degrees of freedom which evolve in time so as to satisfy the constraint of fixed electrostatic potential for each configuration of the electrolyte. The present work clarifies the role of the overall electroneutrality constraint, as well as the link between the averages computed within the Born-Oppenheimer approximation and that of the full constant-potential ensemble. This allows us in particular to derive a complete fluctuation-dissipation relation for the differential capacitance, that includes a contribution from the charge fluctuations (around the charges satisfying the constant-potential and electroneutrality constraints) also present in the absence of an electrolyte. 
We provide a simple expression for this contribution from the elements of the inverse of the matrix defining the quadratic form of the fluctuating charges in the energy. We then illustrate numerically the validity of our results, and recover the expected result for an empty capacitor with structureless electrodes at large inter-electrode distances. By considering a variety of liquids between graphite electrodes, we confirm that this contribution to the total differential capacitance is small compared to that induced by the thermal fluctuations of the electrolyte. 
\end{abstract} 

\maketitle


\section{Introduction}

The electrode-electrolyte interface, where a metallic solid meets an ionic fluid, is at the heart of all electrochemistry and of many applications including sensors~\cite{shao_graphene_2010}, energy storage in batteries~\cite{goodenough_challenges_2010} and supercapacitors~\cite{salanne2016a}, or the harvesting of energy from sunlight~\cite{bak_photo-electrochemical_2002} or salinity gradients in estuaries~\cite{brogioli_extracting_2009}. In order to go beyond the traditional picture of electric double layers in electrolytes or its extensions for ionic liquids (see \textit{e.g.} Refs.~\citenum{parsons_review-doublelayer_1990,fedorov2014a} for reviews), molecular simulation has become an essential tool to investigate the properties of such interfaces, providing both new fundamental understanding~\cite{carrasco2012a,striolo2016a} and insights for practical applications~\cite{merlet2012a,merlet2013d,vatamanu_charge_2017,borodin2017b,simoncelli2018a}.

A description of the electronic structure of the metal requires in principle \textit{ab initio} calculations, usually at the Density Functional Theory level, but their computational cost limits the size of the simulated system on the liquid side, or introduces drastic simplification of its description, \textit{e.g.} using an implicit polarizable model. As a result, classical molecular dynamics has become the standard approach to describe the interfacial liquid and sample its configurations over the relevant length and time scales. This requires in turn an appropriate description at this level of the metal. While treating the electrodes as uniformly charged, structureless or atomically resolved walls has been common practice, several studies underlined the importance of accounting for the polarization of the metal by the interfacial ions and molecules~\cite{merlet2013a,merlet_computer_2013,breitsprecher2015a,haskins_evaluation_2016,vatamanu2017a}. In specific geometries, the continuum picture of image charges can be exploited to compute the electrostatic interactions in simulations of explicit electrolytes between implicit walls~\cite{breitsprecher2015a,girotto_simulations_2017}. 

A more flexible approach, allowing to also capture the atomic structure of the electrode, is to treat the charge of each electrode atom as an additional degree of freedom, which fluctuates in response to the microscopic configuration of the electrolyte in order to maintain the electrode at a constant potential rather than charge. This approach, described in more detail below, was proposed by Siepmann and Sprik to model the tip of a scanning tunneling microscope approaching a planar metal surface\cite{siepmann1995a}, and later adapted by Reed \textit{et al.} to the case of an electrochemical cell\cite{reed2007a}. The use of (Gaussian) charge distributions with fluctuating magnitude to represent the electrode atoms opened the way to the simulation of a variety of electrochemical systems, involving platinum and carbon electrodes and a variety of liquids including molten salts, pure water, aqueous or organic solutions of electrolytes, room temperature ionic liquids and water-in-salts~\cite{reed2008a,willard2009a,willard2013a,limmer_hydration_2013,burt2016a,vatamanu2011a,li_capacitive_2018}. Since then, other approaches have been proposed to account for the polarization of the metal in response to its environment, which share with the Siepmann and Sprik model the treatment of the atomic charges as additional degrees of freedom and the expression of the energy of the system as a quadratic form of these charges, but differ in the coefficient of this form: based on a tight-binding approximation in Ref.~\citenum{pastewka2011a} and on the electronegativity equalization method in Ref.~\citenum{onofrio2015b}.

An important consequence of imposing the potential of the electrodes is that their charge fluctuates. Such charge fluctuations in electrical circuits have been identified as a source of information on the system by Nyquist and Jonhson\cite{johnson1928a,nyquist1928a}, who derived a relation between the differential capacitance and the variance of the charge fluctuations. Charges fluctuations are also exploited to infer information on redox reactions and corrosion processes~\cite{bertocci_noise_1995,cottis_interpretation_2001} or more recently for sensing in nanofluidic channels~\cite{zevenbergen_electrochemical_2009,singh_stochastic_2011,zevenbergen_stochastic_2011}. On the theoretical side, the charge fluctuations in constant-potential simulations have been investigated using \textit{ab initio} simulations~\cite{bonnet2012a} as well as with the primitive model of electrolytes between (fluctuating) uniformly charged smooth electrodes~\cite{kiyohara2007a}. 

In previous work~\cite{limmer_charge_2013}, the charge fluctuations in molecular dynamics simulations of nanocapacitors described by the fluctuating charge model of electrodes was examined. In particular, it was shown how histogram reweighting techniques can be used to sample the properties of nanocapacitors as a function of voltage and to efficiently evaluate the differential capacitance in such simulations. This then allowed to identify a voltage-induced order-disorder transition in the first adsorbed layer of a room temperature ionic liquid on graphite, associated with a peak in differential capacitance~\cite{merlet2014a,rotenberg2015a}, or to study the effects of solution composition on the electrochemical response of a double layer capacitor~\cite{uralcan2016a}. Haskins and Lawson later identified an additional contribution to the capacitance of the system, representing ``the self-capacitance of the electrodes under the influence of the electrolyte structure'', which they found to be small (though non-negligible) and relatively insensitive to the value of voltage for the considered systems~\cite{haskins_evaluation_2016}.

In the present work, we revisit the statistical mechanics of the constant-potential ensemble, for the generic case where the energy is a quadratic form of the fluctuating charges, by introducing explicitly the constraint of global electroneutrality of the system. We investigate the sampling of phase space corresponding to the full constant-potential ensemble, where electrolyte configurations and electrode charges are decoupled, and that achieved in the Born-Oppenheimer approximation, where a single set of electrode charges corresponds to each electrolyte configuration, and discuss the possibility to compute observables with the latter. We obtain explicit analytical results for a number of properties, including the configuration-dependent potential shift previously introduced as a Lagrange multiplier to enforce the electroneutrality constraint, the configuration-dependent set of charges satisfying the constant-potential and electroneutrality constraints and the differential capacitance of the system. We obtain an explicit expression for the term identified as the self-capacitance in Ref.~\citenum{haskins_evaluation_2016}, which we show to be independent not only of voltage but also of the configuration of the electrolyte. This term, which corresponds to the capacitance of the electrolyte-free capacitor, is expressed simply from the elements of the inverse of the matrix defining the quadratic form of the fluctuating charges in the energy. We then illustrate numerically the validity of our results, and recover in particular, at large inter-electrode distance, the expected result for an empty capacitor with structureless electrodes. By considering pure water, an aqueous electrolyte solution and a room temperature ionic liquid between graphite electrodes, we confirm that this contribution to the total differential capacitance is small compared to that induced by the thermal fluctuations of the electrolyte. 
Theoretical results are introduced and derived in Section~\ref{sec:theory}, while applications are considered in Section~\ref{sec:applications}.


\section{Statistical mechanics of the constant-potential ensemble}
\label{sec:theory}

The classical description of metallic electrodes with fluctuating
charges\cite{siepmann1995a,reed2007a} involves mobile ions and molecules of the electrolyte, with $N$ atomic positions $\bfrN$ and momenta ${\bf p}^N$, as well as $M$ immobile electrode atoms with atom-centered charge distributions (such as point charges or Gaussians) with fluctuating magnitudes $\bfq=\{q_1,\dots,q_M\}$. While the following derivations can be extended to the case of mobile electrode atoms (or at least mobile rigid electrodes) by introducing the corresponding degrees of freedom, we will restrict ourselves to the case of fixed electrodes for simplicity. 
The Hamiltonian of the system is written as
\begin{align}
\label{eq:hamiltonian}
\mathH(\bfrN, {\bf p}^N, \bfq) &= 
\mathK({\bf p}^N) + \mathUz(\bfrN) + \mathU_{el}(\bfrN,\bfq)
\; ,
\end{align}
with $\mathK$ the kinetic energy of the electrolyte, and where the potential energy is split into $\mathU_{el}(\bfrN,\bfq)$, the electrostatic interaction involving the fluctuating charges (both between them and with the electrolyte), and $\mathUz(\bfrN)$, which contains all other terms: electrostatic interactions within the electrolyte and all the non-electrostatic terms within the
electrolyte and with the electrode atoms.
The electrostatic term involving the fluctuating charges is quadratic in the charges, written as
\begin{align}
\mathU_{el}(\bfrN,\bfq) &= \frac{\bfqT\matA\bfq}{2} - \bfqT \bfBz
\; ,
\end{align}
where the symmetric $M\times M$ matrix $\matA$ depends on the positions of the electrode atoms
and the parameters describing the charge distribution on each atom,
while the components of the vector $\bfBz$ are the electrostatic potential 
due to the electrolyte on each electrode atom.
Explicit expressions for $\matA$ and $\bfBz$ for the particular case of Gaussian charge distributions,
taking into account two-dimensional Ewald summation for electrostatic interactions, can be found in Refs.~\citenum{reed2007a,gingrich2010a}.
The set of electrostatic potentials on each electrode atom is given by the gradient of $\mathU_{el}$ with respect to $\bfqT$,
\begin{align}
\label{eq:elecpot}
\frac{\partial \mathU_{el}(\bfrN,\bfq)}{\partial \bfqT} &= \matA\bfq - \bfBz
\; ,
\end{align}
and depends on the positions of the electrode atoms and on the electrolyte configuration.

\subsection{Constant-potential ensemble}

We consider the constant-potential ensemble in which the electrostatic
potential on each electrode atom is fixed to a prescribed value.
In addition, the number $N$ of electrolyte atoms, the number $M$
of electrode atoms, the volume $V$ of the system and the temperature $T$
are also considered as fixed in the present work.
We consider a capacitor with two electrodes with potentials $\psi_L$ and $\psi_R$ and corresponding voltage,
\begin{align}
\Dpsi&=\psi_L-\psi_R
\; ,
\end{align} 
with electrode atoms ordered such that the first $M_L$ correspond 
to the left electrode and the remaining $M_R$ to the right electrode. We form the vector of constant potentials
\begin{align}
\label{eq:defDandE}
\Psi^T &= ( \underbrace{\psi_L, \dots, \psi_L}_{M_L\ {\rm atoms}}, 
 \underbrace{\psi_R, \dots, \psi_R}_{M_R\ {\rm atoms}} )
= \Dpsi\bfDT + \bpsi \bfET
\; ,
\end{align}
where we have introduced the average potential
\begin{align}
\bpsi=\frac{M_L\psi_L+M_R\psi_R}{M_L+M_R}
\end{align}
and two constant, orthogonal vectors
\begin{align}
\label{eq:defDandE2}
\bfET &= ( 1, \dots, 1) 
\quad {\rm and} \quad
\bfDT = ( \underbrace{ \alpha_L, \dots, \alpha_L}_{M_L\ {\rm atoms}}, 
 \underbrace{\alpha_R, \dots, \alpha_R}_{M_R \ {\rm atoms}} )
\; ,
\end{align}
with
\begin{align}
\alpha_L &= \frac{M_R}{M_L+M_R}
\quad {\rm and} \quad \alpha_R=\alpha_L-1 = -\frac{M_L}{M_L+M_R}
\; .
\end{align}
For symmetric capacitors, $M_L=M_R$ and the components of $\bfD$ are equal to $\pm\frac{1}{2}$.
The total charge on both electrodes is given by
\begin{align}
Q_{tot} &= Q_L+Q_R =\sum_{i=1}^{M_L} q_i + \sum_{i=M_L+1}^{M_L+M_R} q_i 
= \bfqT \bfE \ .
\end{align}
With this notation, the constraint of global charge electroneutrality (to be discussed in the next section) can be compactly written as
\begin{align}
\label{eq:electroneutrality}
\bfqT \bfE &= \bfET \bfq = 0 \ . 
\end{align}
Then the total charge of the two electrodes are opposite, which is conveniently expressed as
\begin{align}
\label{eq:QLQR}
Q_L &= - Q_R = \bfqT \bfD  = \bfDT \bfq \ . 
\end{align}
Similarly, from Eq.~\ref{eq:electroneutrality}, the product between charges and potentials simplifies to
\begin{align}
\label{eq:QPsi}
\bfPsi^T \bfq &= \bfqT \bfPsi = \Dpsi \bfDT \bfq = Q_L\Dpsi  = - Q_R\Dpsi  \ .
\end{align}
This quantity corresponds to the work exchanged with a charge reservoir
(the external circuit which connects the two electrodes) when charging
the capacitor from $\bfq=0$ to a charge distribution $\bfq$ under the
fixed potentials $\bfPsi$. 

\subsection{Partition function} \label{section:Z}

In the constant-potential ensemble, the charge distribution $\bfq$ on the 
electrodes fluctuates and the corresponding partition function 
is an integral over electrolyte configurations
(we do not consider the fluctuations of the conjugate momenta of the electrolyte, which are statistically independent)
and electrode charge distributions. Two important points must be taken into account
to compute this partition function. The Boltzmann
factor needs to include the work $\bfqT \bfPsi$ in addition to the
Hamiltonian Eq.~\ref{eq:hamiltonian}. 
Additionally, we enforce the
constraint of global electroneutrality, Eq.~\ref{eq:electroneutrality},
since in an experiment, charge is transferred from one electrode to the other, keeping the total charge of the system unchanged. The partition function therefore reads
\begin{align}
\label{eq:Zdef}
\mathZ &= \int\dbfrN\int\dbfq\ e^{ -\beta\left[ \mathUz(\bfrN)
+ \frac{\bfqT\matA\bfq}{2} - \bfqT \bfBz - \bfqT \bfPsi \right]}
\delta( \beta \bfqT \bfE )
\nonumber \\
&= \int\dbfrN e^{ -\beta \mathUz(\bfrN) }
\int\dbfq\ e^{ -\beta\left[ \frac{\bfqT\matA\bfq}{2} - \bfqT (\bfBz +\bfPsi) \right]}
\delta( \beta \bfqT \bfE )
\;,
\end{align}
where $\beta=1/k_BT$, with $k_B$ Boltzmann's constant,
and the Dirac $\delta$ function ensures that only charge distributions satisfying global neutrality are considered. 

In order to perform the constrained Gaussian integral over the charge 
distributions in Eq.~\ref{eq:Zdef}, we introduce the Fourier representation of the Dirac function as
\begin{align}
\delta( \beta \bfqT \bfE )
&= \frac{1}{2\pi} \int_{-\infty}^{+\infty} {\rm d}k\ e^{ik\beta \bfqT \bfE}
\ .
\end{align}
We then reverse the order of the integrations over $\bfq$ and $k$
and perform first the Gaussian integral over $\bfq$:
\begin{align}
\label{eq:Z2}
\mathZ &= 
\int\dbfrN e^{ -\beta \mathUz(\bfrN) }
\frac{1}{2\pi} \int_{-\infty}^{+\infty} {\rm d}k
\int\dbfq\ e^{ -\beta\left[ \frac{\bfqT\matA\bfq}{2} - \bfqT (\bfBz +\bfPsi + ik\bfE) \right]}
\nonumber \\
&= 
\frac{1}{2\pi} \sqrt{ \frac{ (2\pi)^M }{ \beta^M \det \matA} }
\int\dbfrN e^{ -\beta \mathUz(\bfrN) } 
e^{ +\frac{\beta}{2} 
\left[ \bfBz +\bfPsi \right]^T \matAinv \left[ \bfBz +\bfPsi \right]}
\int_{-\infty}^{+\infty} {\rm d}k\ e^{ -\alpha k^2 + \gamma k}
\end{align}
where we have introduced the scalars 
$\alpha = \frac{\beta}{2} \bfET\matAinv\bfE$
and 
$\gamma = i \beta  \bfET\matAinv \left[ \bfBz +\bfPsi \right]$.
Performing the Gaussian integral over $k$ and after some
simple algebra, we can rewrite
\begin{align}
\label{eq:Z3}
\mathZ &= 
K \int\dbfrN e^{ -\beta \mathUz(\bfrN) }
e^{ +\frac{\beta}{2} 
\left[ \bfBz +\bfPsi - \chi(\bfrN)\bfE \right]^T \matAinv \left[ \bfBz +\bfPsi  -
\chi(\bfrN)\bfE \right]}
\end{align}
with a prefactor 
\begin{align}
K=\frac{1}{2\pi} \sqrt{ \frac{ (2\pi)^M }{ \beta^M \det \matA} }
\sqrt{\frac{ 2 \pi}{\beta \bfET\matAinv\bfE}} \; ,
\end{align}
and where we introduced the scalar
\begin{align}
\label{eq:chi}
\chi(\bfrN) &= \frac{1}{\bfET\matAinv\bfE} \bfET\matAinv \left[ \bfBz +\bfPsi \right]
\; .
\end{align}
This quantity can be interpreted as a configuration-dependent potential shift,
since $\bfPsi-\chi\bfE$ simply changes all values of the potentials
by a constant. This additional constant is a Lagrange multiplier enforcing the electroneutrality constraint, but leaving the voltage between the electrodes invariant.
The present derivation provides an explicit expression of this configuration-dependent potential shift $\chi(\bfrN)$ enforcing electroneutrality.

\subsection{Sampling the constant-potential ensemble with Born-Oppenheimer dynamics}

In classical simulations, the constant-potential ensemble can be sampled
using either Monte Carlo simulations or molecular dynamics. Siepmann and Sprik first proposed in Ref.~\citenum{siepmann1995a} to use Car-Parrinello dynamics by introducing fictitious dynamics for the charge degrees of freedom. Reed \textit{et al.} later used Born-Oppenheimer dynamics in Ref.~\citenum{reed2007a}, in which the additional degrees of freedom $\bfq$ are relaxed at each timestep by imposing the constraint of constant-potential on the charges
\begin{align}
\label{eq:fixpot}
\frac{\partial \mathU_{el}(\bfrN,\bfq)}{\partial \bfqT} &= \Psi \; .
\end{align}
Such a Born-Oppenheimer approximation is frequently used in \textit{ab initio} molecular dynamics, which considers the electronic degrees of freedom to instantaneously adapt to the nuclei movement and thus decouples electronic and atomic degrees of freedom.

Without enforcing the electroneutrality constraint (Eq.~\ref{eq:electroneutrality}), the set of charges that satisfies this constant-potential constraint is given by $\matAinv (\bfBz+\Psi)$, as can be immediately seen from Eqs~\ref{eq:elecpot} and~\ref{eq:fixpot}. In Eq. \ref{eq:Z3}, we recognize the set of charges $\bfqs$ satisfying both the constraints of fixed potentials and of global electroneutrality for a given configuration which is given by
\begin{align}
\label{eq:qstar}
\bfqs(\bfrN) &= \matAinv \left[ \bfBz +\bfPsi - \chi(\bfrN)\bfE \right]
= \matS \left[ \bfBz +\bfPsi \right]
\;,
\end{align}
where we have used Eq.~\ref{eq:chi} and defined the symmetric matrix
\begin{align}
\label{eq:defS}
\matS &\equiv \matAinv - \frac{ \matAinv\bfE\bfET\matAinv}{ \bfET\matAinv\bfE }
\;.
\end{align}
We have therefore obtained an explicit expression of the configuration-dependent set of charges satisfying the constraints of fixed potentials and global electroneutrality.
Note that the reference used to define the set of fixed potentials $\bfPsi$ is irrelevant,
as expected. Indeed, shifting all potentials by a constant value 
leads via Eq.~\ref{eq:qstar} to the same set of charges.

From Eq.~\ref{eq:Z3}, the partition function is expressed
using this particular set of charges as
\begin{align}
\label{eq:Z4}
\mathZ &= K
\int\dbfrN e^{ -\beta \mathUz(\bfrN) +\frac{\beta}{2}
[\bfqs(\bfrN)]^T \matA \bfqs(\bfrN)}
\; ,
\end{align}
where we have introduced an equivalent expression
of the electrostatic energy in the Boltzmann factor.

As highlighted earlier, the Born-Oppenheimer dynamics simulations do not sample the constant-potential ensemble. 
Neglecting the fluctuations of the charges around their optimum yields a partition function, 
$\mathZ ^\mathrm{BO}$, which does not include the integral over the additional degrees of freedom $\bfq$, since these are 
now function of the positions $\bfrN$ and replaced by $\bfqs(\bfrN)$. We will write $\bfqs$ in the following but one should bear in mind this explicit dependence on $\bfrN$. 
To write $\mathZ^\mathrm{BO}$, we replace in Eq. \ref{eq:Zdef} the integral over $\bfq$ by enforcing the constraint $\bfq = \bfqs$ for each electrolyte configuration. Using the definition of $\bfqs$, Eq. \ref{eq:qstar}, and noting that $\bfqsT \bfE = 0$, we simplify the Boltzmann factor as
\begin{align}
\label{eq:ZdefBO}
\mathZ^\mathrm{BO} 
&= \int\dbfrN\ e^{ -\beta \mathUz(\bfrN)  +  \frac{\beta}{2} \bfqsT\matA\bfqs} \;.
\end{align}
We can now express the probability of finding a microscopic state in each ensemble as
\begin{align}
\mathP(\bfrN, \bfq) &= \frac{e^{ -\beta\left[ \mathUz(\bfrN) 
+ \frac{\bfqT\matA\bfq}{2} - \bfqT (\bfBz + \bfPsi) \right]} \delta( \beta \bfqT \bfE )}
{K \int\dbfrN e^{ -\beta \mathUz(\bfrN) +\frac{\beta}{2} \bfqsT \matA \bfqs}} \\
\mathP^\mathrm{BO}(\bfrN) &= \frac{e^{ -\beta \mathUz(\bfrN)  +  \frac{\beta}{2} \bfqsT\matA\bfqs}}
{\int\dbfrN e^{ -\beta \mathUz(\bfrN)  +  \frac{\beta}{2} \bfqsT\matA\bfqs}} \; .
\end{align}
Using Eq. \ref{eq:qstar} and simple algebra, it therefore follows that the probability of a given configuration $\{ \bfrN, \bfq \}$ in the constant-potential ensemble
is related to the probability of a given configuration $\{ \bfrN \}$ in the Born-Oppenheimer ensemble as
\begin{align}
\mathP(\bfrN, \bfq) &= \mathP^\mathrm{BO}(\bfrN) K^{-1} \delta( \beta \bfqT \bfE ) e^{ -\frac{\beta}{2} (\bfq - \bfqs)^T\matA(\bfq - \bfqs)} \; .
\end{align}
We then write the probability distribution of an observable $\mathO$ in the constant-potential ensemble as
\begin{align} \label{eq:probO}
\mathp(\mathO) &= \int\dbfrN \int \dbfq \mathP(\bfrN, \bfq) \delta( \mathO(\bfrN, \bfq) - \mathO) \nonumber \\
 &= \int\dbfrN \mathP^\mathrm{BO}(\bfrN) K^{-1} \int \dbfq e^{ -\frac{\beta}{2} (\bfq - \bfqs)^T\matA(\bfq - \bfqs)} \delta( \beta \bfqT \bfE ) \delta( \mathO(\bfrN, \bfq) - \mathO) \; .
\end{align}
For observables that depend exclusively on the positions, \textit{i.e.} $\mathO(\bfrN, \bfq) = \mathO(\bfrN)$, the second Dirac $\delta$ function
can be moved out of the integral over $\bfq$ and, following the same steps as in Section \ref{section:Z}, it is straightforward to show that
the probability distribution in each ensemble are equal, \textit{i.e.} $\mathp(\mathO) = \mathp^\mathrm{BO}(\mathO)$. Therefore, charge-independent observables,
such as density profiles or structure factors, are correctly sampled in Born-Oppenheimer simulations.

We now turn to observables expressed as 
\begin{align} \label{eq:taylorO}
\mathO(\bfrN, \bfq) &= \mathO^*(\bfrN) + (\bfq - \bfqs)^T \bfG^*(\bfrN) \; ,
\end{align}
where $\mathO^*(\bfrN)=\mathO(\bfrN, \bfqs)$ and $\bfG^*(\bfrN) = \nabla_\bfq \mathO | _{\bfqs}$ are the observable and its gradient taken for $\bfq = \bfqs$ 
and depend only on the positions $\bfrN$. Such observables include all linear combinations of the charges, such as the total charge of an electrode, 
and can serve as an approximation for more generic observables as a 
Taylor expansion around $\bfqs$.
To calculate the probability distribution of $\mathO$, we start by introducing the Fourier representation of both Dirac functions in Eq. \ref{eq:probO}, the second one being
\begin{align}
\delta( \mathO(\bfrN, \bfq) - \mathO)
&= \frac{1}{2\pi} \int_{-\infty}^{+\infty} {\rm d}m\ e^{im(\mathO(\bfrN, \bfq) - \mathO)} \, .
\end{align}
Reversing the order of the integrals, inserting Eq. \ref{eq:taylorO} and carrying out the Gaussian integrals over $\bfq$ then $k$ and $m$ results in a compact expression
\begin{align}\label{eq:probO1}
\mathp (\mathO) &= \int \dbfrN \mathP^\mathrm{BO}(\bfrN) \sqrt{\frac{\beta}{2 \pi [\bfG^*(\bfrN)]^T \matS \bfG^*(\bfrN)}} \ e^{- \frac{\beta}{2} \frac{[ \mathO^*(\bfrN) - \mathO]^2}{[\bfG^*(\bfrN)]^T\matS\bfG^*(\bfrN)}} \, ,
\end{align}
where we used the definition of $\matS$ in Eq. \ref{eq:defS}. 
This formulation shows that the Born-Oppenheimer sampling of the observable $\mathO$ lacks a contribution due to the charge degrees of freedom
that is expressed here as Gaussian fluctuations of this observable around a mean value $\mathO^*(\bfrN)$ with a variance $\beta ^{-1} [\bfG^*(\bfrN)]^T \matS \bfG^*(\bfrN)$.

As an example, which will be used in the next section, we apply Eq. \ref{eq:probO1} to the total charge on the left electrode, $Q_L(\bfrN) = \bfqT \bfD = Q^*_L(\bfrN) + (\bfq - \bfqs)^T \bfD$. In this particular case, the gradient $\bfG^*(\bfrN) = \bfD$ 
is independent of the electrolyte positions $\bfrN$.  
The average of $Q_L$ over the full phase space corresponding to the constant-potential is easily computed as the Gaussian integral
\begin{align} \label{eq:avgQ}
\avgpsi{Q_L} &= \int  {\rm d}Q_L \; \mathp(Q_L) \; Q_L
\nonumber \\
&= \int \dbfrN \mathP^\mathrm{BO}(\bfrN) \sqrt{\frac{\beta}{2 \pi \bfD^T \matS \bfD}} \int {\rm d}Q_L \; Q_L \ e^{- \frac{\beta}{2} \frac{(Q^*_L - Q_L)^2}{\bfD^T\matS\bfD}} \nonumber \\
&= \int \dbfrN \mathP^\mathrm{BO}(\bfrN) Q_L^*(\bfrN) = \avgBO{Q_L^*}
\end{align}
and therefore coincides with the average of $Q_L^*$ sampled in the Born-Oppenheimer approximation.
Similarly, the average of $Q_L^2$ is given by the second moment of the distribution, which is the sum of the squared mean and the variance,
\begin{align} \label{eq:avgQ2}
\avgpsi{Q_L^2} &= \avgBO{Q_L^{*2}} + \beta ^{-1} \avgBO{\bfDT \matS \bfD} 
\nonumber \\
&= \avgBO{Q_L^{*2}} + \beta ^{-1} \bfDT \matS \bfD \; ,
\end{align}
where for the second equality we have used the fact that
$\bfDT \matS \bfD$ does not depend on the electrolyte configuration $\bfrN$.
Therefore, this second moment does not simply coincide with the average of $Q_L^{*2}$ sampled in the Born-Oppenheimer approximation. It contains 
an additional contribution from the fluctuations of the charge around $\bfqs$ for each configuration of the electrolyte. 
This result applies to any observable that depends linearly on the charges and proves that Born-Oppenheimer dynamics sample correctly their average
value but lacks a term in their fluctuations, that we can express as
a function of the gradient of the observable with respect to the charges.


\subsection{Charge and capacitance}

We now turn to the calculation of the differential capacitance which is defined as the derivative of the
average charge of the electrode $\avgpsi{Q_L}$
with respect to voltage,
\begin{align}
\label{eq:Cdiff1}
C_\mathrm{diff} &= \frac{\partial \avgpsi{Q_L}}{\partial\Dpsi}
\; .
\end{align}
In order to evaluate the latter, we will use the fact that
the charge and differential capacitance are the first
and second order derivatives of the free energy 
$F=-\beta ^{-1} \ln\mathZ$
with respect to voltage $\Dpsi$.
From the expression of $\mathZ$ given in Eq.\ref{eq:Zdef},
and using Eq.~\ref{eq:QPsi},
the first derivative of the free energy is easily computed as
\begin{align}
\label{eq:1stDeriv}
-\frac{1}{\beta}\frac{\partial \ln \mathZ}{\partial\Dpsi}
&= - \frac{1}{\beta \mathZ}\int\dbfrN e^{ -\beta \mathUz(\bfrN) }
\int\dbfq\ (\beta \bfqT \bfD) e^{ -\beta\left[ \frac{\bfqT\matA\bfq}{2} - \bfqT (\bfBz +\bfPsi) \right]}
\delta( \beta \bfqT \bfE )
\nonumber \\
&= - \avgpsi{\bfqT \bfD} = - \avgpsi{Q_L}
\end{align}
where the average is taken over microscopic configurations
in the constant-potential ensemble.
The free energy change during the charge of the capacitor,
initially at zero voltage, to finite voltage $\Dpsi_{max}$,
is therefore
\begin{align}
\label{eq:freeencharge}
\Delta F &= -\int_0^{\Dpsi_{max}} \avgpsi{Q_L} {\rm d}\Dpsi
\ = \int_0^{\Dpsi_{max}} \avgpsi{Q_R} {\rm d}\Dpsi 
\; ,
\end{align}
the reversible electrical work exchanged with
the charge reservoir. 
The differential capacitance is then obtained from the second
derivative of the free energy with respect to voltage as
\begin{align}
\label{eq:2ndDeriv}
C_\mathrm{diff} &= \frac{ \partial  \avgpsi{Q_L}}{ \partial \Dpsi} 
= \beta \avgpsi{\delta Q_L^2} \; ,
\end{align}
with $\delta Q_L=Q_L-\avgpsi{Q_L}$.
The differential capacitance $C_\mathrm{diff}$ thus corresponds to the fluctuations of the total charge on one electrode
in the constant-potential ensemble.
To make the link with the observables computed with Born-Oppenheimer dynamics, we introduce Eq. \ref{eq:avgQ} and \ref{eq:avgQ2}:
\begin{align}
\label{eq:Cdiff2}
C_\mathrm{diff} &= \beta \avgpsi{Q_L^2} - \beta  \avgpsi{Q_L}^2 
\nonumber \\
&= \beta \avgBO{Q_L^{*2}} + \bfDT \matS \bfD - \beta  \left(\avgBO{Q_L^{*}} \right)^2 \nonumber \\
&= \beta \avgBO{\delta Q_L^{*2}} + \bfDT \matS \bfD
\; .
\end{align}

This is the main practical result of the present work. Compared to the result of Ref.~\citenum{limmer_charge_2013},
where the last term was missing, and to Ref.~\citenum{haskins_evaluation_2016},
the present derivation shows how an extra capacitance emerges from the Gaussian fluctuations of the charges, which have been integrated
out in the Born-Oppenheimer ensemble of configurations where only the charge distribution $\bfqs$
satisfying the constant-potential and electroneutrality constraints is considered explicitly.

Interestingly, given Eq. \ref{eq:avgQ}, an alternative derivation starting from the Born-Oppenheimer partition function Eq. \ref{eq:ZdefBO} and considering 
$C_\mathrm{diff} = \partial \avgBO{Q_L^*} / \partial\Dpsi$, the additional term in the differential capacitance
appears as $\avgBO{ \partial Q_L^*/\partial\Dpsi}$.
Indeed, using Eq.~\ref{eq:QLQR} and $\frac{\partial\bfqs}{\partial\Dpsi}=\matS\bfD$, we obtain the equality:
\begin{align}
\avgBO{\frac{\partial Q_L^*}{\partial\Dpsi}}
&= \avgBO{\bfDT \frac{\partial\bfqs(\bfrN)}{\partial\Dpsi}}
= \avgBO{ \bfDT \matS\bfD } = \bfDT \matS\bfD 
\, .
\end{align}

This allows to further interpret the extra capacitance $\bfDT \matS \bfD$ as the average response of the electrode to voltage for each microscopic configuration
of the electrolyte, which was mentioned by Haskins and Lawson in Ref.~\citenum{haskins_evaluation_2016}.
The present derivation further shows that it is independent of the microscopic
configuration of the electrolyte as well as voltage, and provides an explicit 
expression from the $\matA$ matrix describing the electrostatic interactions
between all electrode atoms. This corresponds to the capacitance
in the absence of any electrolyte,
\begin{align}
\label{eq:Cdiffempty}
C_\mathrm{diff}^\mathrm{empty} &= \bfDT \matS\bfD 
\; .
\end{align}
It combines additively with the contribution arising from charge fluctuations induced by the thermal fluctuations of the electrolyte,
\begin{align}
\label{eq:Celectrlolyte}
C_\mathrm{diff}^\mathrm{electrolyte} &= \beta \left[ \avgBO{ Q_L^{*2}} -
\left( \avgBO{Q_L^*}\right)^2 \right] 
= \beta \avgBO{\delta Q_L^{*2}}
\; , 
\end{align}
so that Eq.~\ref{eq:Cdiff2} corresponds to an equivalent
circuit of capacitors in parallel.

A particularly simple expression in terms of the elements of the $\matAinv$ matrix
can be obtained, if one introduces the following block decomposition:
\begin{align}
\label{eq:defblocks}
\matAinv &= \left(
\begin{array}{c|c}
(\matAinv)_{LL} & (\matAinv)_{LR} \\
\hline
(\matAinv)_{RL} & (\matAinv)_{RR}
\end{array}
\right)
\; , 
\end{align}
with diagonal blocks of size $M_L\times M_L$ and $M_R\times M_R$
and off-diagonal blocks of size $M_L\times M_R$ and $M_R\times M_L$.
Using simple matrix algebra, together with the definitions
Eq.~\ref{eq:defS} of $\matS$ and  
Eq.~\ref{eq:defDandE2} of $\bfD$ and $\bfE$,
one can show that the extra capacitance corresponding
to the empty capacitor is given by 
\begin{align}
\label{eq:DTSDfromAinv}
C_\mathrm{diff}^\mathrm{empty} &= \bfDT \matS\bfD =
\frac{ \Sigma_{LL}\Sigma_{RR} - \Sigma_{LR}^2 }{ \Sigma_{LL} + \Sigma_{RR} + 2\Sigma_{LR} }
\; , 
\end{align}
where $\Sigma_{LL}$, $\Sigma_{RR}$ and $\Sigma_{LR}$ are the sums of the elements
of the blocks defined in Eq.~\ref{eq:defblocks}. In the particular but common case of identical electrodes, with $\alpha_L=-\alpha_R=1/2$ and $\Sigma_{LL}=\Sigma_{RR}$, one also has the simple result $C_\mathrm{diff}=\bfDT \matAinv \bfD=(\Sigma_{LL}-\Sigma_{LR})/2$.
In general, Eq.~\ref{eq:DTSDfromAinv} provides a simple route to the calculation of the capacitance of arbitrarily shaped electrodes in a continuum dielectric, which could be used in the design of nano-structured capacitors\cite{yu_hybrid_2013}.


\section{Applications}

\label{sec:applications}

The analytical result Eq.~\ref{eq:Cdiffempty} can be used to compute the 
differential capacitance of the empty capacitor, which also contributes
to the differential capacitance in the presence of an electrolyte
(see Eq.~\ref{eq:Cdiff2}), if the matrix $\matA$ can be inverted.
In the limit of large systems, \textit{i.e.} large number of electrode atoms 
and large distance between electrodes compared to their atomic structure, 
one should recover the result for an empty parallel plate
capacitor with structureless electrodes, 
namely a capacitance per unit area $C_\mathrm{diff}/\mathcal{A}=\varepsilon_0/L$,
with $\varepsilon_0$ the vacuum permittivity and $L$ the distance between the
electrodes. Performing the matrix inversion explicitly becomes cumbersome when the number of atoms is large, so that it is difficult to take such a limit analytically. Nevertheless, we do recover it numerically, as shown below." 

\begin{figure}[ht!]
\centering
  \includegraphics[width=0.45\textwidth]{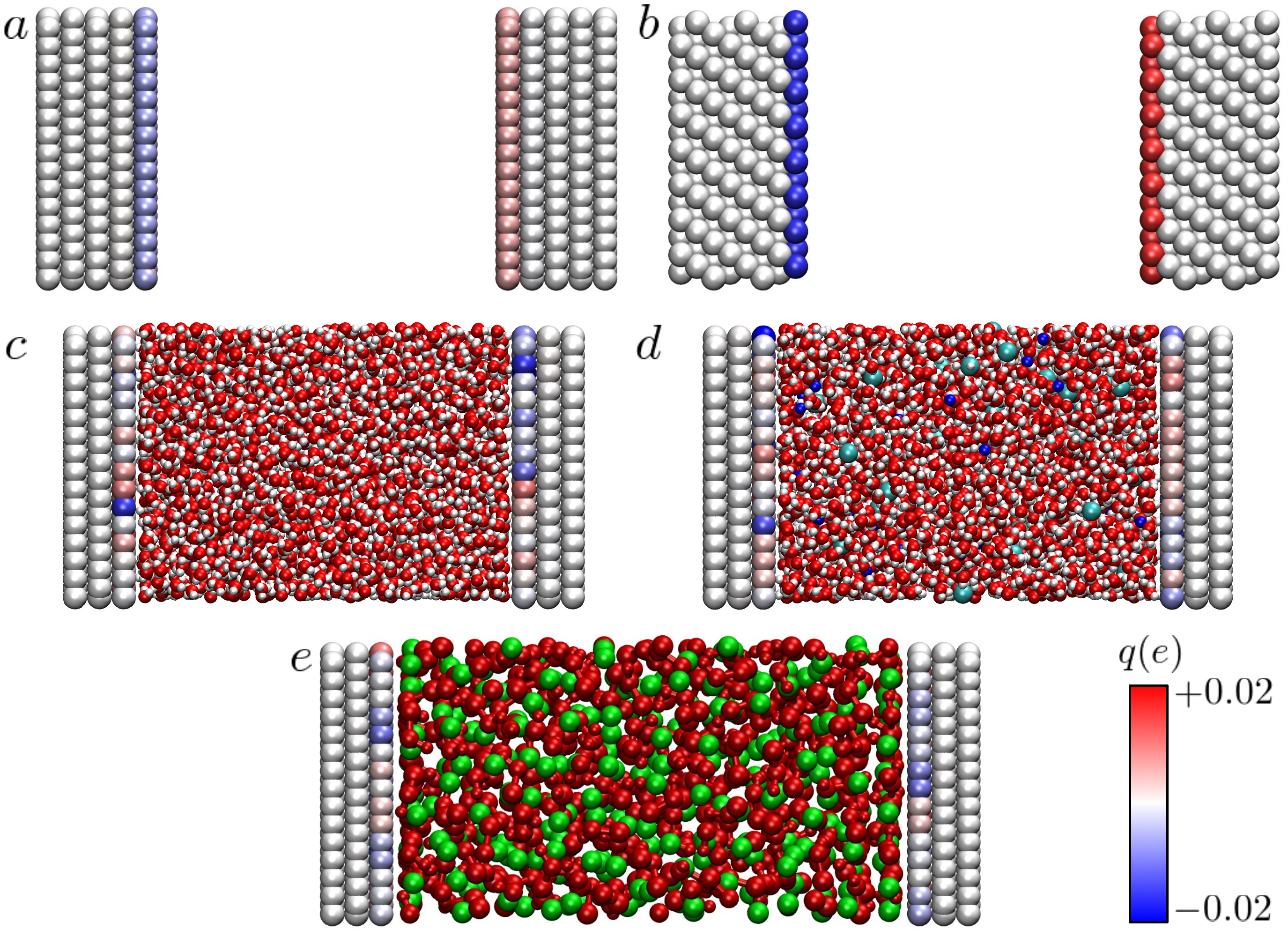}
  \caption{
	Snapshots of the systems: 
	(a) empty graphite capacitor
	(b) empty platinum capacitor
	(c) graphite capacitor with pure water
	(d) graphite capacitor with 1M NaCl aqueous electrolyte
	(e) graphite capacitor with BMI-PF$_6$ ionic liquid.
	The color scale indicates the charge of the electrode atoms for
	the considered configurations, at $\Delta\psi=20$~V and 0~V
	for the empty and full capacitors, respectively.
	}  
  \label{fig:systems}
\end{figure}

\subsection{Systems and simulation details}

In the present work, we consider several capacitors, illustrated in Figure~\ref{fig:systems}: 
two empty capacitors, consisting of either graphite (panel a) or platinum (panel b) electrodes,
and three graphite capacitors filled with pure water (panel c),
a 1M NaCl aqueous electrolyte (panel d), or a room temperature ionic liquid,
namely 1-butyl-3-methylimidazolium hexafluorophosphate (BMI-PF$_6$, panel e).
We use the molecular dynamics code MetalWalls to run constant potential simulations with 
potential differences $\Delta \psi$ 
ranging from 0.0 to 20.0 V. Each electrode atom has a Gaussian charge distribution of width $\eta^{-1}=0.55$~\AA. The charges $\bfqs$ satisfying the constant-potential constraint are determined at each time step using conjugate gradient minimization.

Two-dimensional boundary conditions are used with no periodicity in the $z$
direction and we use an accurate 2D Ewald summation method to compute electrostatic interactions~\cite{reed2007a,gingrich2010a}. 
We use a cutoff of 17~\AA\ for both the short range part of the Coulomb interactions and the 
intermolecular interactions. For the latter we use the truncated shifted Lennard-Jones potential.
For graphite electrodes, the lateral box dimensions are $34.101\times 36.915$~\AA$^2$ with 
5 atomic planes (\textit{i.e.} 2400 carbon atoms) per electrode. For platinum, we consider 
the ($111$) surface of the {\it fcc} crystal structure with a lattice parameter $a=3.9$~\AA; 
the lateral box dimensions are $41.550\times 36.000$~\AA$^2$ with 8 atomic planes (\textit{i.e.} 1800 platinum atoms)
per electrode. 
For the filled capacitors, the electrolytes are composed of 2160 SPC/E water molecules, 2160 SPC/E water molecules and 39 Na$^+$-Cl$^-$ ion pairs, and 240 BMI$^+$-PF$_6^-$ ion pairs, respectively. Force field parameters can be found in references \citenum{berendsen1987a, werder_watercarbon_2003,cole1983a} for water, \citenum{berendsen1987a, dang_mechanism_1995, werder_watercarbon_2003,cole1983a} for aqueous NaCl and \citenum{roy2010a, cole1983a} for BMI-PF$_6$. The simulation boxes are equilibrated at constant atmospheric pressure for 500~ps by applying a constant pressure force to the electrodes then the electrodes separation is fixed to the equilibrium value $L =$ 55.11~\AA, 56.2~\AA, and 74.11~\AA, respectively for water, aqueous NaCl and BMI-PF$_6$. The aqueous systems were run at 298~K with a timestep of 1~fs whereas the ionic liquid at 400~K with a timestep of 2~fs, using a chain Nos\'e-Hoover thermostat. Each system is run for at least 4~ns to extract the total charge fluctuations on the electrodes.

\subsection{Electrolyte-free capacitors}

\begin{figure}[ht!]
\centering
  \includegraphics[width=0.4\textwidth]{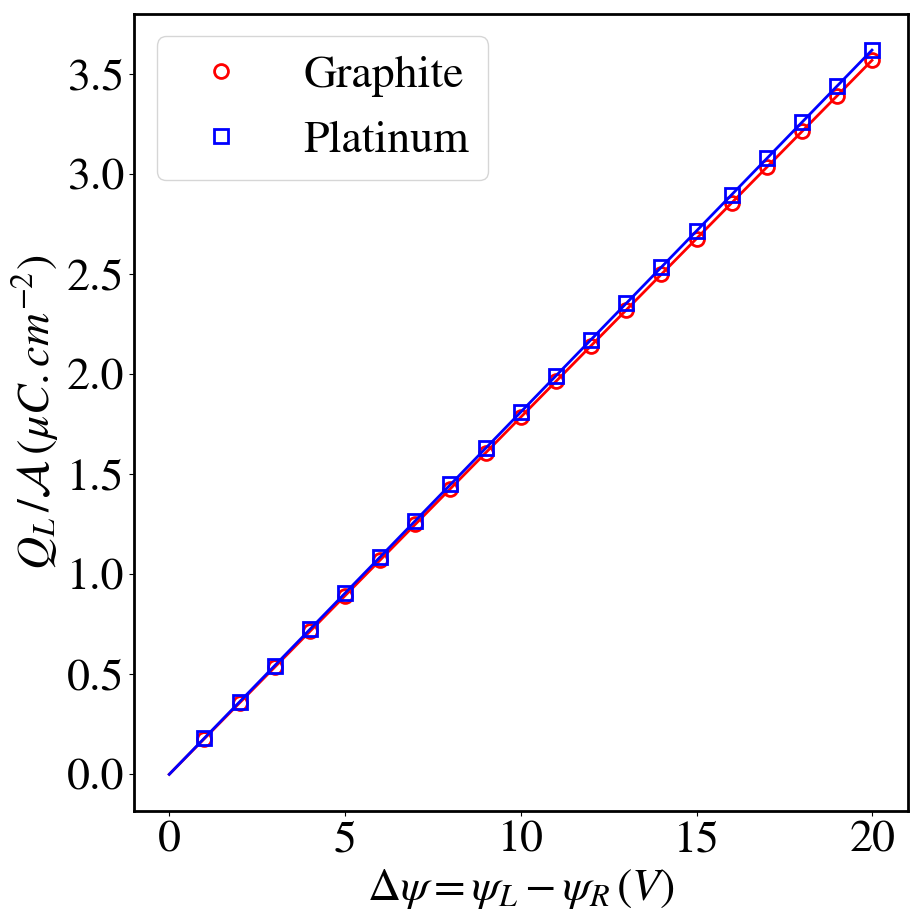}
  \caption{
	Charge of the left electrode $Q_L$ as a function of the applied
	voltage $\Dpsi=\psi_L-\psi_R$, for the two empty graphite
	or platinum capacitors shown in Figs.~\ref{fig:systems}a and b,
	respectively (see text for details).
    The distance between the first atomic planes of the electrodes is $L=50.0$~\AA.
	The symbols indicate the charge determined numerically for various
	applied voltages between $1.0$ and $20.0$~V, while the lines are the predictions of
	Eq.~\ref{eq:Cdiffempty}.}
  \label{fig:emptyQvsPsi}
\end{figure}

The contribution of the empty capacitor
can be computed from the positions of the electrode atoms
and the width of the Gaussian charge distributions (which define the elements
of the matrix $\matA$) via Eq.~\ref{eq:Cdiffempty}.
This result does not depend on voltage $\Dpsi$, so that
the charge of the empty capacitor should be proportional to $\Dpsi$.
Figure~\ref{fig:emptyQvsPsi} compares this prediction for
two empty capacitors, consisting of either graphite or platinum electrodes, 
with the charge determined numerically for several voltages.
The numerical results confirm both the linear response,
which means that the integral capacitance, $Q_L/\Dpsi$ (ratio between the electrode
charge and applied voltage), does not depend on voltage and 
is equal to the differential capacitance, $C_\mathrm{diff}$,
and that the slope is correctly predicted by Eq.~\ref{eq:Cdiffempty}
from the positions of the electrode atoms and the width of the Gaussian
charge distributions.

\begin{figure}[ht!]
\centering
  \includegraphics[width=0.4\textwidth]{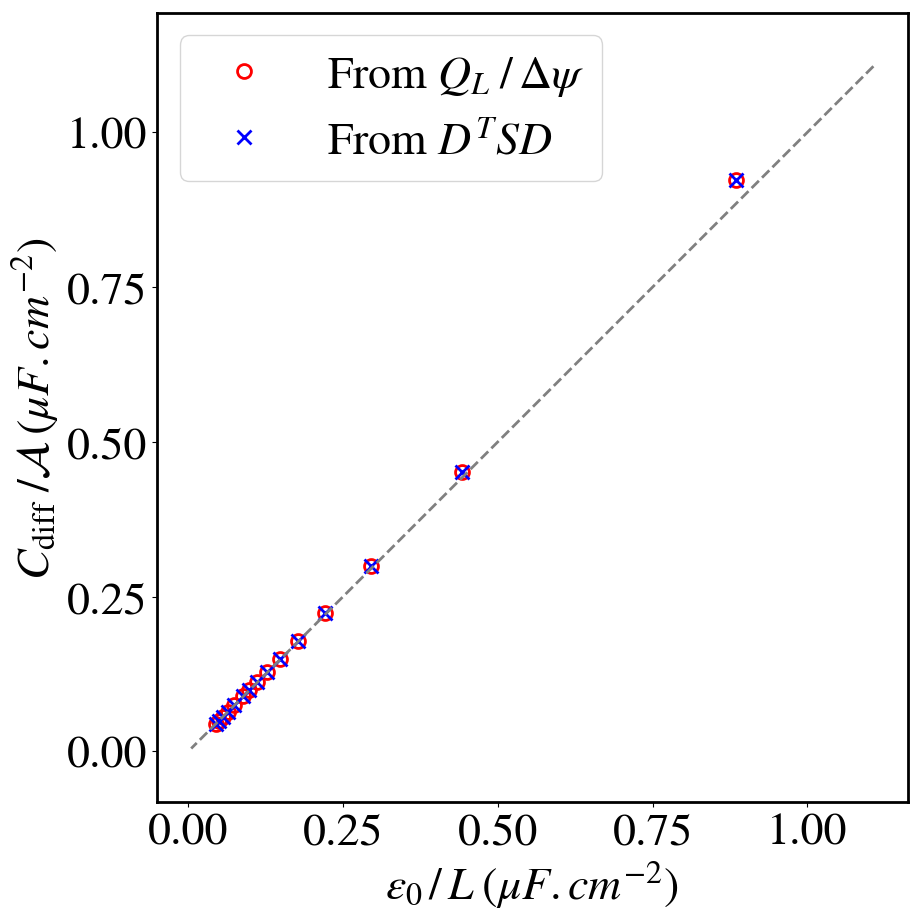}
  \caption{
	Differential capacitance per unit area, $C_\mathrm{diff}/\mathcal{A}$, as a function of the analytical prediction 
	for an empty capacitor with structureless electrodes:
	$\varepsilon_0 / L$, with $\mathcal{A}$ the area and $L$ the
	distance between the electrodes ranging from $10$ to $200$~\AA, for the empty capacitors
	(see Fig.~\ref{fig:systems}a and b). Red circles correspond to the slope
	of Fig.~\ref{fig:emptyQvsPsi}, while the blue crosses are the predictions from
	Eq.~\ref{eq:Cdiffempty}.
	The dashed line corresponds to $y=x$.
	}  
  \label{fig:emptyCdiff}
\end{figure}

Figure~\ref{fig:emptyCdiff} then shows the capacitance per unit area,
$C_\mathrm{diff}/\mathcal{A}$, as a function of the analytical result for
structureless electrodes, $\varepsilon_0 / L$, for the empty graphite capacitor,
with distances $L$ ranging between $10$ and $200$~\AA.
The numerical results for the atomically resolved electrodes converge to
the analytical result for large inter-electrode distances, as expected.
Surprizingly, the difference is already small ($< 5$\%)
for a distance $L=10$~\AA, which is not much larger than
the distance between graphite planes ($3.354$~\AA)
or even the interatomic distance within planes ($1.42$~\AA).
The numerical results are in fact even better described 
as $\varepsilon_0 / L_{eff}$, with an effective inter-electrode
distance $L_{eff}=L-d$, with $d=0.4$~\AA\ comparable to the Gaussian width $\eta^{-1}$ (not shown).

\subsection{Full capacitors}

\begin{figure}[ht!]
\centering
  \includegraphics[width=0.4\textwidth]{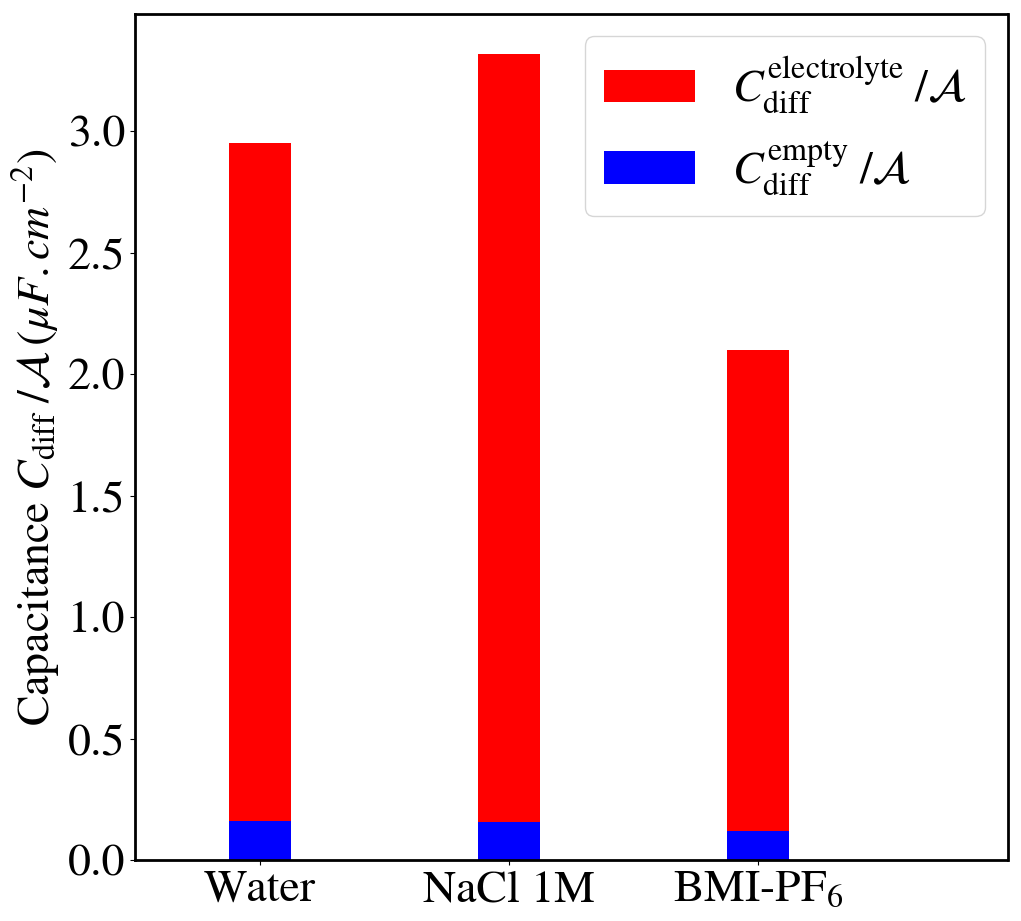}
  \caption{
	Contributions to the differential capacitance per unit area $C_\mathrm{diff}/\mathcal{A}$ 
	(see Eq.~\ref{eq:Cdiff2}): from the empty capacitor,
	$C_\mathrm{diff}^\mathrm{empty}=\bfDT \matS\bfD$ 
	and from the polarization fluctuations induced by the electrolyte,
	$C_\mathrm{diff}^\mathrm{electrolyte}=\beta\avgBO{\delta Q_L^{*2}}$.
	Results are given for the three capacitors shown in panels c, d and e
	of Figure~\ref{fig:systems}.
	}  
  \label{fig:fullCdiff}
\end{figure}

We finally examine the relative contributions to the total differential capacitance
in the presence of a liquid between the electrodes (see Eq.~\ref{eq:Cdiff2}):
that corresponding to the empty capacitor, $C_\mathrm{diff}^\mathrm{empty}=\bfDT \matS\bfD$,
and that due to the thermal fluctuations of the electrolyte,
$C_\mathrm{diff}^\mathrm{electrolyte}=\beta \avgBO{\delta Q_L^{*2}}$.
The results are shown on Figure~\ref{fig:fullCdiff} for the graphite 
capacitors containing either water, an 1M NaCl aqueous electrolyte 
or BMI-PF$_{6}$ (see panels c, d and e of Figure~\ref{fig:systems}).
In all cases, the contribution derived in this work, $C_\mathrm{diff}^\mathrm{empty}$,
is small (between 5 and 6\%) compared to $C_\mathrm{diff}^\mathrm{electrolyte}$
already obtained in Ref.~\citenum{limmer_charge_2013}.
Such an observation is consistent with that of Haskins and 
Lawson~\cite{haskins_evaluation_2016} and explains why this additional
contribution may have been forgotten in earlier work.
For the case of pure water, we note that the contribution of the liquid
is not simply $\varepsilon_0 (\varepsilon_r-1)/ L$, with $\varepsilon_r$
the relative permittivity, because of the structuration of water at
the interface\cite{willard2009a,bonthuis_profile_2012,jeanmairet_study_2019}.


\section{Conclusions}

We revisited the statistical mechanics of the constant-potential ensemble, for the generic case where the energy is a quadratic form of the fluctuating charges, by introducing explicitly the constraint of global electroneutrality of the system. We obtained explicit analytical results for a number of properties, including the configuration-dependent potential shift previously introduced as a Lagrange multiplier to enforce the electroneutrality constraint, as well as the configuration-dependent set of charges satisfying the constant-potential and electroneutrality constraints. 
We investigated the sampling of phase space corresponding to the full constant-potential ensemble and that achieved in the Born-Oppenheimer approximation, 
and discussed the possibility to compute observables with the latter. In particular, we showed that for linear combination of atomic charges, 
the average value is correctly sampled, while the fluctuations need to be corrected by a term for which we obtained an explicit expression. These results justify the use of Born-Oppenheimer dynamics to sample the constant-potential ensemble. 

We derived an analytical expression of the differential capacitance of the system including an additional self-capacitance, as identified in Ref.~\citenum{haskins_evaluation_2016}, which we showed to be independent of voltage and of the configuration of the electrolyte. This term, which corresponds to the capacitance of the electrolyte-free capacitor, is expressible from the elements of the inverse of the matrix defining the quadratic form of the fluctuating charges in the energy. 
We illustrated numerically the validity of our results, and recovered in particular, at large inter-electrode distance, the expected result for an empty capacitor with structureless electrodes ($\varepsilon_0\mathcal{A}/L$). By considering a variety of liquids between graphite electrodes, 
we confirmed that this contribution to the total differential capacitance is small compared to that induced by the thermal fluctuations of the electrolyte. 

The present work applies to all classical electrode models with a Hamiltonian which can be expressed as a quadratic form of fluctuating charges. In particular, it is readily applicable to the semiclassical Thomas-Fermi model recently proposed to tune the metallicity of electrodes in molecular simulations~\cite{dufils_semiclassical_2019}, in order to interpret intriguing observations in Atomic Force Microscopy experiments on ionic liquids~\cite{comtet2017a,kaiser2017a}. It would be interesting to extend the present work to open systems, in which the exchange of electrolyte with a reservoir results in fluctuations in the number of ions and solvent molecules, which play an important role in the context of blue energy harvesting~\cite{vanroij_statistical_2014}. Other directions include the simulation of reactive systems~\cite{dwelle_constant_2019} for electrochemical applications, as well as the coupling of electrical and mechanical properties, including forces induced by charge fluctuations~\cite{drosdoff_charge-induced_2016} or to analyze recent Surface Force Balance experiments~\cite{perez-martinez_surface_2019}.

\section*{Conflicts of interest}
There are no conflicts to declare.

\section*{Acknowledgements}
BR and DTL were supported by the France-Berkeley Fund from the University of 
California, Berkeley. 
BR acknowledges financial support from the French Agence Nationale de la Recherche 
(ANR) under Grant No. ANR-17-CE09-0046-02 (NEPTUNE). 
This project has received funding from the European Research Council (ERC) under
the European Union's Horizon 2020 research and innovation programme (grant
agreement No. 771294).



\providecommand*{\mcitethebibliography}{\thebibliography}
\csname @ifundefined\endcsname{endmcitethebibliography}
{\let\endmcitethebibliography\endthebibliography}{}

\end{document}